\documentstyle[epsfig]{elsartme}

\newcommand{\nc}{\newcommand}
\nc{\rnc}{\renewcommand}
\nc{\acs}{\arraycolsep}
\nc{\mc}{\multicolumn}
\nc{\bsk}{\baselineskip}
\nc{\vsp}{\vspace}
\nc{\hsp}{\hspace}
\nc{\stl}{\setlength}
\nc{\stc}{\setcounter}
\nc{\addl}{\addtolength}
\nc{\beq}{\begin{equation}}
\nc{\eeq}{\end{equation}}
\nc{\beqa}{\begin{eqnarray}}
\nc{\eeqa}{\end{eqnarray}}
\nc{\NL}{\newline}
\nc{\tfrac}[2]{\raisebox{.4ex}{\tiny $\frac{#1}{#2}$}} 
\nc{\romlist}{ \setcounter{num1}{0}%
  \begin{list}{(\roman{num1})}{\usecounter{num1}} }
\nc{\arblist}{ \setcounter{num1}{0}%
  \begin{list}{(\arabic{num1})}{\usecounter{num1}} }
\nc{\alphlist}{ \setcounter{num2}{0}%
  \begin{list}{(\alph{num2})}{\usecounter{num2}} }
\nc{\bullist}{\begin{list}{$\bullet$}{ }}
\nc{\nr}{\\ \hline}
\nc{\hrl}{{\center \stl{\unitlength}{\textwidth} 
 \begin{picture}(1,0)  \put(0,0){\line(1,0){1}}
 \end{picture} \vsp{.001\bsk} }}
\nc{\cents}{{\scriptsize$\mbox{\rm C}\!\!\!\mbox{\raisebox{.2ex}%
{$|$}}\,\,\,\,$}}
\nc{\figsp}[5]{\begin{figure}[#1] \vsp{#2} \caption[#4]{#3} 
\label{#5} \vsp{2\bsk} \end{figure}}
\nc{\fig}{\figsp{tbp}}
\nc{\figb}{\figsp{b}}
\nc{\figh}{\figsp{h}}
\nc{\llist}{\begin{list}{}{} \stl{\labelsep}{.4in}}
\nc{\lit}[2]{
 \item[\raggedright #1]{#2}}
\nc{\lbit}[2]{ 
 \item[\raggedright\bf #1]{#2}}
\nc{\lemit}[2]{ 
 \item[\raggedright\em #1]{#2}}
\nc{\lbemit}[2]{ 
 \item[\raggedright\bf\em #1]{#2}}
\nc{\clst}[1]{\stl{\coltwo}{\textwidth}
\addl{\coltwo}{-#1} \addl{\coltwo}{-5.56ex} \newline
\begin{tabular}{p{#1}p{\coltwo}} \citem{}{}}
 
\nc{\citem}[2]{{\raggedright \bf #1} & #2 \\ }
\nc{\cemitem}[2]{{\raggedright \em #1} & #2 \\ }
\nc{\cbemitem}[2]{{\raggedright \bf \em #1} & #2 \\ }
\nc{\cend}{\citem{}{} \end{tabular} 
\mbox{}
} 
\nc{\SSP}{{\rm \hsp{.4in}}}
\nc{\SSPP}{{\rm \hsp{.2in}}}
\nc{\ds}{\displaystyle}
\nc{\tx}{\textstyle}
\nc{\scst}{\scriptstyle}
\nc{\sscst}{\scriptscriptstyle}
\nc{\prt}{\partial}
\nc{\fr}{\frac}
\nc{\lf}{\left}
\nc{\rt}{\right}
\nc{\la}{\langle}
\nc{\ra}{\rangle}
\nc{\V}{\vec}
\nc{\str}{\stackrel}
\nc{\ovl}{\overline}
\nc{\ul}{\underline}
\nc{\ovb}{\overbrace}
\nc{\ub}{\underbrace}
\nc{\wh}{\widehat}
\nc{\B}{\bar}
\nc{\D}{\dot}
\nc{\C}{\cdot}
\nc{\dd}{\ddot}
\nc{\tl}{\tilde}
\nc{\ha}{\hat}
\nc{\nn}{\nonumber}
\nc{\app}{\approx}
\nc{\al}{\alpha}
\nc{\RA}{\rightarrow}
\nc{\LRA}{\leftrightarrow}
\nc{\SRA}{\SSP\rightarrow\SSP}
\nc{\SSRA}{\SSPP\rightarrow\SSPP}
\nc{\dg}{\dagger}
\nc{\vp}{\varphi}
\nc{\ve}{\varepsilon}
\nc{\Dl}{\Delta}
\nc{\dl}{\delta}
\nc{\gm}{\gamma}
\nc{\Gm}{\Gamma}
\nc{\ep}{\epsilon}
\nc{\sg}{\sigma}
\nc{\Sg}{\Sigma}
\nc{\ua}{\uparrow}
\nc{\da}{\downarrow}
\nc{\lam}{\lambda}
\nc{\eql}[1]{\parbox{#1\textwidth}}
\nc{\eqm}[1]{\makebox[#1\textwidth][l]}
\nc{\enu}[1]{\mbox{\hspace{.4in}(\theequation.#1)}}
\nc{\son}{\\ \\ \ds}
\nc{\stw}{\\ & \\ \ds}		%   Equation formatting   %
\nc{\sth}{\\ & & \\ \ds}
\nc{\sfo}{\\ & & & \\ \ds}
\nc{\sfi}{\\ & & & & \\ \ds}
\nc{\A}{& \ds}
\nc{\bbr}{\lf\{\rule[-1.5ex]{0in}{0.01in}\rt.}
\nc{\hf}{\fr{1}{2}}
\nc{\mhf}{\mbox{\footnotesize$\hf$}}
\nc{\dv}{\/!}
\nc{\dint}{\int\!\!\int}
\nc{\tint}{\int\!\!\dint}               % integrals %
\nc{\qint}{\int\!\!\tint}
\nc{\Pd}[2]{\fr{\prt #1}{\prt #2}}
\nc{\Pdt}[1]{\Pd{#1}{t}}
\nc{\Pdx}[1]{\Pd{#1}{x}}
\nc{\Pdy}[1]{\Pd{#1}{y}}
\nc{\Pdz}[1]{\Pd{#1}{z}}           	%  Derivatives  %           
\nc{\Pdr}[1]{\Pd{#1}{r}}
\nc{\Pds}[1]{\Pd{#1}{s}}
\nc{\Dv}[2]{\fr{d#1}{d#2}}
\nc{\Dvt}[1]{\Dv{#1}{t}}
\nc{\Dvx}[1]{\Dv{#1}{x}}
\nc{\Dvy}[1]{\Dv{#1}{y}}
\nc{\Dvz}[1]{\Dv{#1}{z}}
\nc{\Dvr}[1]{\Dv{#1}{r}}
\nc{\Drs}[1]{\Dv{#1}{s}}
\nc{\inpp}[3]{\la #1| #2| #3\ra}
\nc{\inp}[2]{\inpp{#1}{#2}{#1}}
\nc{\rb}[1]{| #1\ra}
\nc{\lb}[1]{\la#1|}
\nc{\dtpp}[2]{\lb{#1}\rb{#2}}		%  Inner Products  %
\nc{\dtp}[1]{\dtpp{#1}{#1}}
\nc{\otpp}[2]{\rb{#1}\lb{#2}}
\nc{\otp}[1]{\otpp{#1}{#1}}
\newcounter{num1} \newcounter{num2}  %for \romlist and \alphlist
\newlength{\coltwo}
\rnc{\L}{{\cal L}}                      %  Miscellaneous  %
\nc{\lapp}{\mbox{\raisebox{-.6ex}{$\,\stackrel{\textstyle <}{\sim}\,$}}}
\nc{\gapp}{\mbox{\raisebox{-.6ex}{$\,\stackrel{\textstyle >}{\sim}\,$}}}
\nc{\als}{\fr{\al_s(Q^2)}{2\pi}}
\nc{\gpx}{g_1^p(x,Q^2)}
\nc{\gpz}{g_1^p(z,Q^2)}
\nc{\muq}{\lf(\fr{\mu^2}{Q^2}\rt)}
\nc{\xy}{(\fr{x}{y})}
\nc{\ASQ}{\al_s(Q^2)}
\nc{\Li}{{\rm Li}_2}
\nc{\dqx}{\Dl q_i(x,Q^2)} \nc{\dqy}{\Dl q_i(y,Q^2)} 
\nc{\dQ}{\Dl q_i(Q^2)}
\nc{\dgx}{\Dl g(x,Q^2)} \nc{\dgy}{\Dl g(y,Q^2)} 
\nc{\dG}{\Dl g(Q^2)}
\nc{\xq}{(x,Q^2)} \nc{\yq}{(y,Q^2)}
\nc{\Tt}{\tl{t}} \nc{\Ts}{\tl{s}} \nc{\Tu}{\tl{u}}
\nc{\Hs}{\ha{s}} \nc{\Ht}{\ha{t}} \nc{\Hu}{\ha{u}}
\nc{\Hsg}{\hat{\sg}}
\nc{\GeV}{\mbox{\rm GeV}}
\nc{\sS}{\!\not{\!s}}  
\nc{\soS}{\!\not{\!s}_1}  \nc{\swS}{\!\not{\!s}_2}
\nc{\pS}{\!\not{\!p}}  \nc{\kS}{\!\not{\!k}}
\nc{\poS}{\!\not{\!p}_1}  \nc{\pwS}{\!\not{\!p}_2}
\nc{\ptS}{\!\not{\!p}_3}  \nc{\pfS}{\!\not{\!p}_4}
\nc{\AS}{\!\not{\!\!A}}  \nc{\ASS}{\!\not{\!\!A}^*}
\nc{\BS}{\!\not{\!\!B}}  \nc{\BSS}{\!\not{\!\!B}^*}
\nc{\Tr}{\mbox{\rm Tr}}
\nc{\pT}{p_T}
\nc{\xT}{x_T}
\nc{\AoS}{\!\not{\!\!A}_1}  \nc{\AoSS}{\!\not{\!\!A}_1^*}
\nc{\AwS}{\!\not{\!\!A}_2}  \nc{\AwSS}{\!\not{\!\!A}_2^*}
\nc{\BoS}{\!\not{\!\!B}_1}  \nc{\BoSS}{\!\not{\!\!B}_1^*}
\nc{\BwS}{\!\not{\!\!B}_2}  \nc{\BwSS}{\!\not{\!\!B}_2^*}
\nc{\aS}{\!\not{\!a}}  \nc{\bS}{\!\not{\!b}}
\nc{\aoS}{\!\not{\!a}_1}  \nc{\boS}{\!\not{\!b}_1}
\nc{\awS}{\!\not{\!a}_2}  \nc{\bwS}{\!\not{\!b}_2}
\nc{\anS}{\!\not{\!a}_n}  \nc{\bnS}{\!\not{\!b}_n}
\nc{\anpoS}{\!\not{\!a}_{n+1}}  \nc{\bnpoS}{\!\not{\!b}_{n+1}}
\nc{\anmoS}{\!\not{\!a}_{n-1}}  \nc{\bnmoS}{\!\not{\!b}_{n-1}}
\nc{\refi}[1]{$^{\,\mbox{\scriptsize \ref{#1}}}$}
\nc{\refii}[2]{$^{\,\mbox{\scriptsize \ref{#1},\ref{#2}}}$}
\nc{\refiii}[3]{$^{\,\mbox{\scriptsize \ref{#1},\ref{#2},\ref{#3}}}$}
\nc{\refr}[2]{$^{\,\mbox{\scriptsize \ref{#1}--\ref{#2}}}$}
\nc{\sint}{\int \!\!}
\nc{\qB}{\stackrel{(-)}{q}}
\nc{\Lam}{\Lambda}

\stl{\textwidth}{6.5in}
\stl{\evensidemargin}{0in}
\stl{\oddsidemargin}{0in}

\begin{document}

%\stl{\bsk}{1.65\bsk}

%\runauthor{Kamal}
\begin{frontmatter}

\title{Transversity and Mass Singularities in
Dimensional Regularization}
\author{Basim Kamal\thanksref{addr} }
\address{Physics Department, Brookhaven National Laboratory, Upton, NY 11973}

\thanks[addr]{Present address: Department of Physics, 3302 Herzberg 
Laboratories, 1125 Colonel By Drive, Ottawa, ON K1S 5B6, Canada.
E-mail: bkamal@physics.carleton.ca\, .}

\begin{abstract}

\begin{parbox}{6.3in}
{\leftskip -.15in %\stl{\bsk}{1.65\bsk}
The proton's transversity distribution will be measured at BNL's
Relativistic Heavy Ion Collider in upcoming experiments using
the transverse Drell-Yan process. Understanding the one-loop
corrections is therefore important. Here,
the collinear structure in transverse Drell-Yan  is
investigated in detail using dimensional regularization and the 
correct behaviour is found, although the mechanism is non-trivial.
The resulting $n$-dimensional transversity splitting function
(and consequently the one-loop transversity distribution and its
two-loop evolution) is found to be the same in both the 
anticommuting-$\gm_5$ scheme and the HVBM scheme. Alternative schemes
are considered.

\vsp{.5cm}

}
\end{parbox}

\end{abstract}
\end{frontmatter}

%\typeout{SET RUN AUTHOR to \@runauthor}

\mbox{\hfill}

\newpage

\section{Introduction}

\vsp{-.6cm}
Recently, two-loop calculations \cite{Kumano,V,Others} of the transversity
splitting function $\Dl_T P_{qq}$ \cite{Artru} relevant to the evolution of 
the transversity distribution $\Dl_T f_q$ (or $h_1^q$) \cite{Ralston} 
(to be measured in upcoming experiments at BNL's Relativistic Heavy
Ion Collider) have been
carried out using dimensional regularization (DREG). Of some concern is the
fact that an anticommuting $\gm_5$ was used in the above determinations.
This is known to be mathematically inconsistent. Fortunately \cite{V} the
traces only involve even numbers of $\gm_5$'s, so we do not anticipate
any inconsistencies for this specific determination.  
Of greater concern is whether DREG
itself is suitable for the calculation of higher order corrections to 
processes with transverse polarization, either in the anticommuting-$\gm_5$ 
scheme or the mathematically consistent 
't Hooft-Veltman-Breitenlohner-Maison (HVBM) \cite{HV,BM} scheme. 
More precisely, it is necessary to verify the correct behaviour of the
squared amplitude, relevant to some subprocess involving transverse 
polarization,
under collinear gluon bremsstrahlung. That behaviour
must be consistent with that of
other regularization schemes in order to yield
a meaningful and process independent $\Dl_T f_q$.
This is the issue we shall address in this paper.

\vsp{-.6cm}

\section{Generalities}

\vsp{-.6cm}

\begin{figure}
\centerline{\epsfig{file=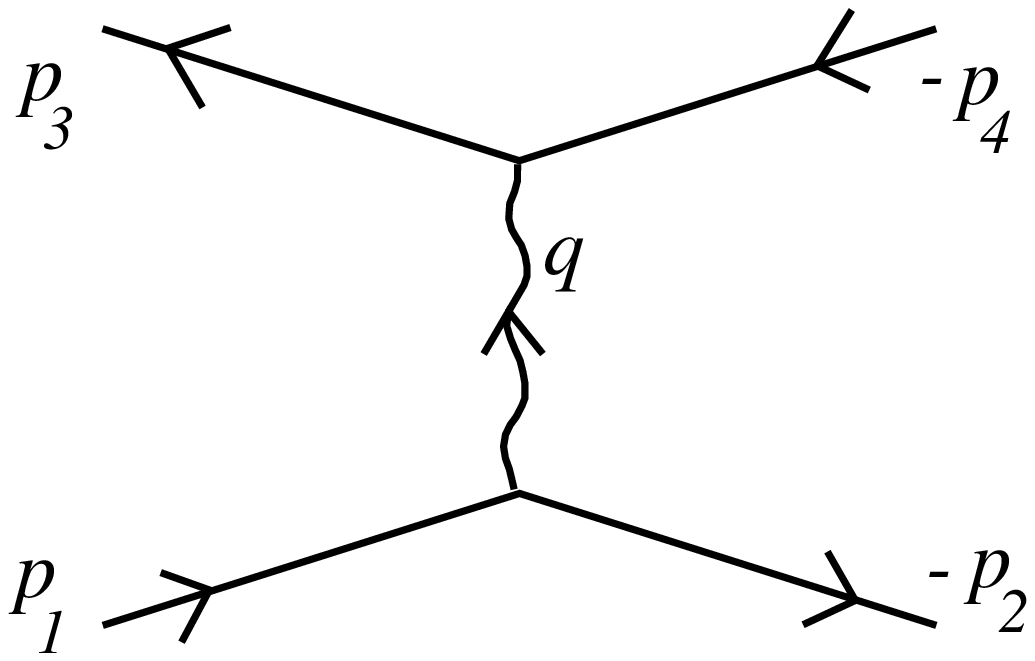,height=4.5cm,width=6.5cm}}
\vspace{10pt}
\caption{Leading order contribution to transverse Drell-Yan.}
\label{Fig1}
\end{figure}

\begin{figure}
\centerline{\epsfig{file=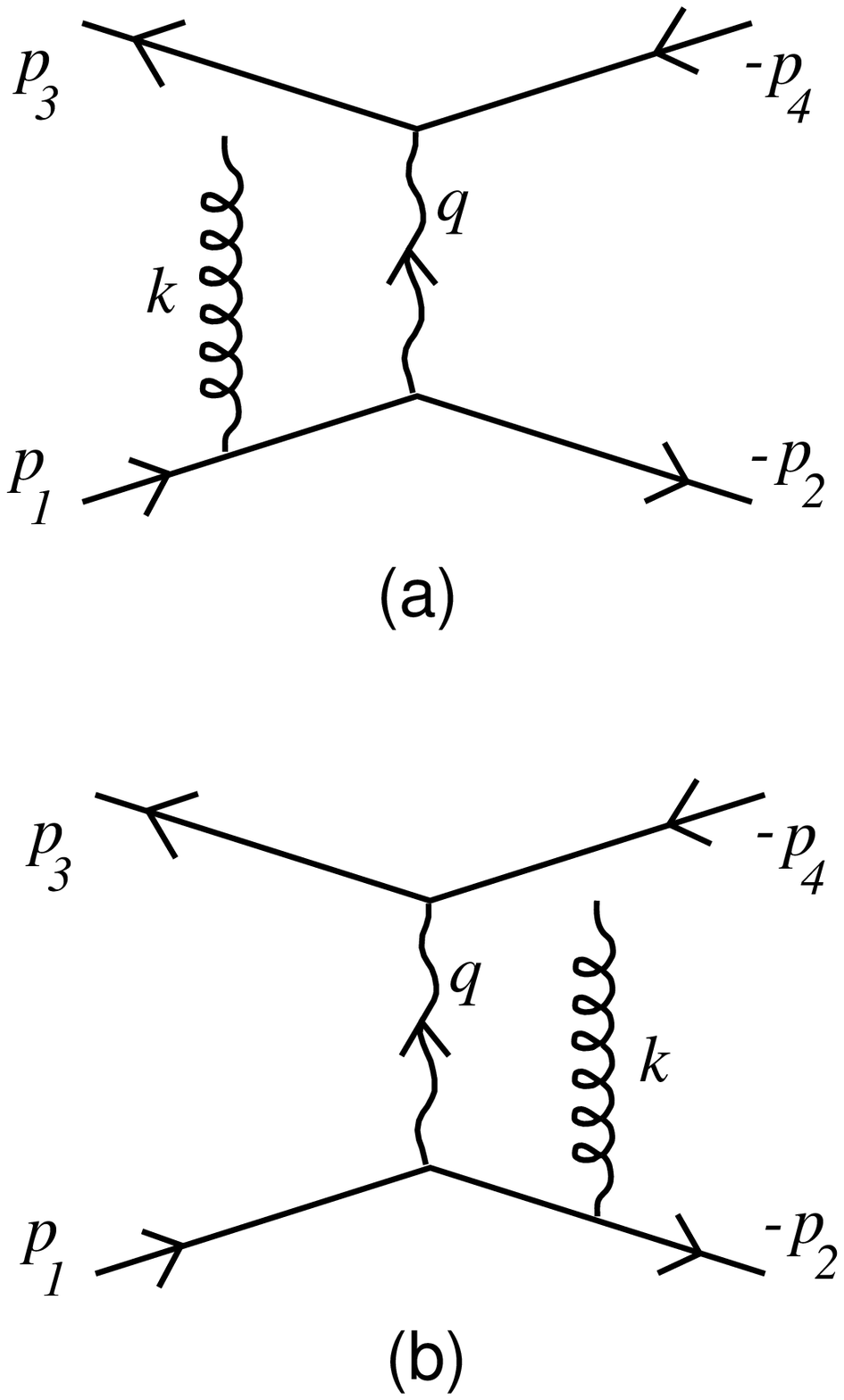,height=11cm,width=6.5cm}}
\vspace{10pt}
\caption{Bremsstrahlung contributions to transverse Drell-Yan.}
\label{Fig2}
\end{figure}

The process we shall consider is {\em transverse Drell-Yan}, where we have
\beq
A(P_1,S_1) + B(P_2,\pm S_2) \RA l^-(p_3) + l^+(p_4) + X.
\eeq
Here $A,B$ denote hadrons with momenta $P_1,P_2$ and transverse spin
vectors $S_1,S_2$. The spin vectors satisfy
%\beq
$S_i\C P_j = 0$, $S_i^2 = -1$.
%\eeq
In leading order, this process is mediated by
\beq
q(p_1,s_1) + \B{q}(p_2,\pm s_2) \RA \gm^*(q) \RA l^-(p_3) + l^+(p_4) \,\,
+ \,\, (q \LRA \B{q}),
\eeq
shown in Fig.\ \ref{Fig1}. In the parton model $p_1 = x_1 P_1$,
$p_2 = x_2 P_2$ for some $0 < x_1, x_2 < 1$.
At next-to-leading order in QCD, we have the gluon loop corrections to 
the above subprocess and the bremsstrahlung subprocess(es)
\beq
q(p_1,s_1) + \B{q}(p_2,\pm s_2) \RA \gm^*(q) + g(k) \RA l^-(p_3) + l^+(p_4)
+ g(k) \,\, + \,\, (q \LRA \B{q}),
\eeq
shown in Fig.\ \ref{Fig2}.
There is no $qg$ subprocess contributing to transverse Drell-Yan. For
the extraction of $\Dl_T f_q$, one is interested in the transversely
polarized  cross section and the corresponding subprocess cross section
defined by
\beq
\Dl_T \sg \equiv \fr{1}{2}[\sg(S_1,S_2) - \sg(S_1,-S_2)], \SSP
\Dl_T \ha{\sg} \equiv \fr{1}{2}[\ha{\sg}(s_1,s_2) - \ha{\sg}(s_1,-s_2)],
\eeq
respectively,
with $s_i=S_i$. We see then that the $S_i$ ($s_i$) represent the ``up''
directions, or spin quantization axes. $\Dl_T \sg$ is obtained by
appropriately convoluting the $\Dl_T f_q(x_i)$ with $\Dl_T \ha{\sg}$.
Comparison with experiment then yields information on $\Dl_T f_q$.
We may define the invariants
\beq \label{invar}
M^2  \equiv  q^2 \equiv sw , \SSPP s \equiv  (p_1+p_2)^2,  \SSPP
t\equiv   (p_1-p_3)^2, \SSPP u\equiv (p_2-p_3)^2.
\eeq
If we let the component
of $p_3$ transverse to the beam axis ($\ha{z}$) define the 
$\ha{x}$ axis, then
in $n=4-2\ve$ dimensions,  in the $p_1,p_2$ c.m., the momenta and
spin vectors are decomposed as
\beqa \nn \label{mompar}
s_{1,2} &=& (0;0,\ldots,0,\ovb{\sin\phi_{1,2}}^y,
\ovb{\cos\phi_{1,2}}^x, \ovb{0}^z), \\ \nn
p_3 &=& |\vec{p}_3|(1;0,\ldots,0,\sin\theta_3,\cos\theta_3), \SSPP
p_{1,2} = \fr{\sqrt{s}}{2} (1;0,\ldots,0,\pm 1), \\
k &=& |\vec{k}| (1; \ldots, \sin\theta_{1,k} \sin\theta_{2,k}
\cos\theta_{3,k}, \sin\theta_{1,k} \cos\theta_{2,k}, \cos\theta_{1,k}),
\eeqa
where, for fixed $M^2$ and $p_3$ direction (but not magnitude),
\beqa \nn \label{p3mag}
|\vec{p}_3| &=& \fr{M^2}{2} \fr{1}{[q_0+|\vec{k}|
(\sin\theta_3\sin\theta_{1,k} \cos\theta_{2,k} + \cos\theta_3
\cos\theta_{1,k})] }, \\ 
|\vec{k}| &=& \fr{\sqrt{s}}{2} (1-w), \SSP
q_0 = \fr{\sqrt{s}}{2} (1+w).
\eeqa
In $k$, the $\ldots$ represent the $n-4$
components to be (trivially) integrated
over.

To obtain a nonvanishing result, we must not integrate fully
over the azimuthal angle of $p_3$, $\phi_3$. Writing
%\beq
$d^{n-1}p_3 = |\vec{p}_3|^{n-2} d |\vec{p}_3|
d^{n-2} \Omega_3$,
%\eeq
we may present the relevant $2\RA 3$ phase space,
\beqa \label{phasesp}
\fr{\Dl_T d \ha{\sg}}{dM^2 d^{n-2}\Omega_3} &=&
\fr{1}{(2\pi)^{5-4\ve}} \fr{M^{2-4\ve}}{2^{2-2\ve}}
\fr{|\vec{k}|^{1-2\ve}}{s^{3/2}} \\ \nn 
& & \times \int\!\! d^{n-2} \Omega_k
\fr{\Dl_T |M|^2}{[q_0+|\vec{k}|
(\sin\theta_3\sin\theta_{1,k} \cos\theta_{2,k} + \cos\theta_3
\cos\theta_{1,k})]^{2-2\ve} },
\eeqa
where $|M|^2$ is the color averaged $2\RA 3$ particle squared amplitude.
In general \cite{Mar}
\beq \label{nint}
\int\!\! d^{m-2} \Omega = 
\int_0^\pi d \theta_1 \sin^{m-3} \theta_1
\int_0^\pi d \theta_2 \sin^{m-4} \theta_2
\ldots
\int_0^\pi d \theta_{m-3} \sin \theta_{m-3}
\int_0^{2\pi} d \theta_{m-2}.
\eeq
The $2\RA 2$ particle phase space is simply
\beq
\fr{\Dl_T d \ha{\sg}}{dM^2 d^{n-2}\Omega_3} =
\fr{2^{2\ve} M^{-4-2\ve}}{(2\pi)^{2-2\ve}} \dl(1-w) \Dl_T |M|^2_{2\RA 2}.
\eeq

Note, our squared amplitude normalization corresponds to the convention
$\sum_\lambda u(p,\lambda) \B{u}(p,\lambda) = \pS/2$ for $p^2=0$, with
$\lambda$ denoting helicity.
Once the above differential cross section contributions are 
obtained and the
virtual contributions and factorization counterterms are added, 
yielding a finite result in the limit $\ve\RA 0$, one can integrate
over $\theta_3$ and obtain $\Dl_T d \ha{\sg}/dM^2 d\phi_3$. The original
calculation of the latter quantity,
at one-loop, was done in \cite{VW} using the 
regularization
scheme where the gluon is given a mass in order to control the
collinear and soft singularities. 
That scheme had already been used successfully in unpolarized
Drell-Yan \cite{Kub}.
Then, in \cite{CKM} the calculation was done
using regularization by dimensional reduction (DRED) \cite{Sie1}. 
The general approach for converting from DRED to DREG, for arbitrary
polarization, was given in \cite{BK1}. That approach requires knowledge
of the $n$-dimensional transversity splitting function, and a more 
careful derivation of it was necessary. This was done
in \cite{V}, where the results of \cite{VW}
were converted to DREG. 
Unfortunately, the $n$-dimensional transversity splitting function
was {\em explicitly} derived in \cite{V} only in the context of a two-loop
calculation.  No details regarding the collinear limit were given, other
than that the correct form resulted. 
The result of \cite{V} for the transverse Drell-Yan
cross section confirms the earlier DRED result of \cite{CKM} and it
confirms the general form of the cross section given in \cite{BK1},
valid for all consistent $n$-dimensional regularization schemes -- 
the scheme dependence is correctly parametrized in terms of the 
$n$-dimensional splitting function.

\vsp{-.6cm}

\section{The squared amplitude}

\vsp{-.6cm}

Let us write for the $2\RA 3$ particle amplitude
%\beq
$M = M_1 + M_2$,
%\eeq
where $M_1,M_2$ are shown in Figures \ref{Fig2}(a),(b) respectively. Then
\beqa \label{trace} \nn
\Dl_T |M|^2 &=& \Dl_T |M_1|^2 + \Dl_T |M_2|^2 
 + 2 \Dl_T M_1 M_2^* \\ \nn
&=& - C \Biggl\{ \fr{1}{[(p_1-k)^2]^2} {\rm Tr}
[\gm_5 \swS \pwS \gm_\al (\poS-\kS) \gm_\mu \gm_5 \soS \poS \gm_\nu
(\poS-\kS) \gm_\beta] \\ \nn
& &   + \fr{1}{[(p_2-k)^2]^2} {\rm Tr}
[\gm_5 \swS \pwS \gm_\mu (\pwS-\kS) \gm_\al \gm_5 \soS \poS \gm_\beta
(\pwS-\kS) \gm_\nu] \\ \nn
& & - \fr{2}{(p_1-k)^2(p_2-k)^2} {\rm Tr}
[\gm_5 \swS \pwS \gm_\al (\poS-\kS) \gm_\mu \gm_5 \soS \poS \gm_\beta
(\pwS-\kS) \gm_\nu] \Biggr\} \\
&& \times g^{\mu\nu} \lf (p_3^\al p_4^\beta + p_4^\al p_3^\beta - \fr{M^2}{2}
g^{\al\beta}\rt ),
\eeqa
where $C$ is an overall factor. Let $\Dl_T |M^n_B|^2$ denote the
$n$-dimensional Born term in DREG. Writing
%\beq
$\Dl_T |M^n_B|^2 = \Dl_T |M^4_B|^2 + \ve \Dl_T |M^\ve_B|^2$, we have
%\eeq
in the anticommuting-$\gm_5$ scheme,
\beqa 
\Dl_T |M^n_B|^2_{\rm AC} &= & C_B \lf[ 2 M^2 p_3\C s_1 \, p_3\C s_2
+ tu s_1\C s_2 + \ve \fr{M^4}{2} s_1\C s_2 \rt] \\
&= & C_B \lf[ \fr{M^4}{4} \sin^2\theta_3 \cos(\phi_1+\phi_2-2\phi_3)
- \ve \fr{M^4}{2} \cos(\phi_1-\phi_2) \rt],
\eeqa
where $C_B$ is an overall factor and clearly $w=1$.
We have now rotated back to a frame with an arbitrary $\ha{x}$ axis
direction via
%\beq
$\phi_{1,2} \RA \tl{\phi}_{1,2} = \phi_{1,2} - \phi_3$. 
%\eeq
In the HVBM scheme
\beq
\Dl_T |M^\ve_B|^2_{\rm HVBM} = - \Dl_T |M^\ve_B|^2_{\rm AC}.
\eeq
At this point, we notice that the Born term has the wrong azimuthal
dependence in DREG. Since the effect is order $\ve$, it will not
manifest itself as long as the Born term factors properly. Then
the $\ve$-dependent part will cancel along with the singularities that
multiply it. This is the case for the virtual corrections and the
soft corrections pose no problems thanks to the Bloch-Nordsieck
mechanism. The remaining question, therefore, is: what is the behaviour
in the limit of collinear gluon bremsstrahlung?

\vsp{-.6cm}

\section{The collinear limit}

\vsp{-.6cm}

In order to investigate the collinear limit, $k \parallel p_1$,
we define the quantities 
\beq
\Dl_T P^{<,4}_{qq}(z) \equiv C_F \lf( \fr{2}{1-z} -2 \rt), \SSP
\Dl_T \tl{P}_{qq}(z) \equiv -C_F(1-z),
\eeq
where the superscript $<$ indicates that $z < 1$. Let us also define
\beq
t'\equiv (p_1'-p_3')^2, \SSPP
u'\equiv (p_2'-p_3')^2, \SSPP
p_1'\equiv w p_1, \SSPP p_2'\equiv  p_2.
\eeq
$p_3'$ is the value which $p_3$ takes in the above
collinear limit, for
fixed $\Omega_3$ and $M^2$. Hence the primed quantities are those
relevant for the kinematics of the Born term which should factor
out in the above limit.
Now, in the collinear limit, $k \parallel p_1$, the $2\RA 3$ particle
squared amplitude in the anticommuting-$\gm_5$ scheme takes the form
(with $w < 1$)
\beqa \nn  \label{coll}
\lefteqn{\Dl_T |M|^2_{\rm coll, AC} =} \\ \nn
&& \fr{4C}{w p_1\C k} \lf[  \fr{\Dl_T |M^n_B|^2_{\rm AC}(t=t')
}{C_B C_F}
[ \Dl_T P^{<,4}_{qq}(w) + \ve \Dl_T \tl{P}_{qq}(w)] \rt. \\ 
&& \lf. - (\ve w tu + \fr{\ve^2 M^4}{2}) \fr{k\C s_1 k \C s_2}{p_1\C k} 
+ \ve M^2 p_3\C s_2 \fr{k\C s_1(wt+u+M^2)}{p_1\C k}
\rt],
\eeqa
plus terms which are nonsingular and do not give rise to scheme
dependences -- the $O(\ve^2)$ terms above may also be dropped.
In the HVBM scheme, we find
\beq
\Dl_T |M|^2_{\rm coll, HVBM} = \Dl_T |M|^2_{\rm coll, AC} 
(\ve \LRA - \ve) + O(\ve^2,\ve \ha{k}^2),
\eeq
so that the structure is identical in both schemes, with only
the sign of $\ve$ reversed (including in the Born term). There are no
finite contributions from $\ha{k}^2$ integrations, where $\ha{k}$
is the vector whose only nonzero components are
the components of $k$ having index greater than three (the
timelike index being zero) in $n$ dimensions; they
contribute like an extra power of $\ve$. Also, none of the $O(\ve^2,
\ve \ha{k}^2)$ terms dropped multiply a soft divergent term.

The term $\sim k\C s_1\, k\C s_2$ in (\ref{coll})
depends on the ``azimuthal'' angles
$\theta_{2,k}$ and $\theta_{3,k}$, of the gluon, in $n$ dimensions.
In the collinear limit, $k \parallel p_1$ (i.e.\ $\theta_{1,k}\RA 0$)
those angles must be integrated over since they are unconstrained. 
From (\ref{p3mag}),(\ref{phasesp}), we see that
the azimuthal dependent part in the denominator of the phase space 
and that of $t,u$
vanishes in the collinear limit by order $\sin\theta_{1,k}$
(see below),
so that the only finite dependence of that term on
$\theta_{2,k}, \theta_{3,k}$ comes from the factor 
$\sim k\C s_1\, k\C s_2$.
This term gives a finite contribution due to the $1/p_1\C k$ factor. 
Since it is multiplied by
$\ve$, it only gives a finite contribution to the cross section from the 
phase space region $\theta_{1,k}\RA 0$, where the $1/\ve$ pole coming
from integrating the overall  $1/p_1\C k$ factor arises. Hence, we may
perform the azimuthal integration in that limit. From
(\ref{mompar}),(\ref{phasesp}),(\ref{nint}) we see that 
we may make the effective substitution
\beqa \label{subst} \nn
\fr{k\C s_1 k \C s_2}{p_1\C k^{\,2}} &\RA & \fr{1}{p_1\C k} \lf[
\int_0^\pi d \theta_{2,k} \sin^{-2\ve} \theta_{2,k}
\int_0^\pi d \theta_{3,k} \sin^{-1-2\ve} \theta_{3,k}
\fr{k\C s_1 k \C s_2}{p_1\C k} \rt] \\ \nn
&& / 
\lf[
\int_0^\pi d \theta_{2,k} \sin^{-2\ve} \theta_{2,k}
\int_0^\pi d \theta_{3,k} \sin^{-1-2\ve} \theta_{3,k}
\,\, 1 \rt] \\ \nn
&=& (1-w) \fr{\cos(\phi_1-\phi_2)}{(1-\ve) p_1\C k}
= - (1-w) \fr{s_1\C s_2}{1-\ve} \fr{1}{p_1\C k} \\
&=&  - (1-w) \fr{s_1\C s_2}{p_1\C k} + O(\ve).
\eeqa
Note, it would make no sense to retain the $O(\ve)$ term since our approach
only determines the finite contribution.
Similarly, the last term of (\ref{coll}) picks up an additional finite
azimuthal dependence from the azimuthal dependence of $t$ and $u$,
which is of order $\sin\theta_{1,k}$ and can be obtained by series
expanding (\ref{p3mag}) about $\sin\theta_{1,k}=0$.
Without this extra dependence, the term 
would not contribute. The procedure is then similar to that used for
the term $\sim k\C s_1\, k\C s_2$.
We end up with the effective
substitution
\beq \label{ssub}
\fr{k\C s_1(wt + u + M^2)}{p_1\C k^{\,2}}
\RA \fr{2(1-w) s_1\C p_3}{p_1\C k} + O(\ve).
\eeq
Substituting (\ref{subst}) and (\ref{ssub}) in (\ref{coll}), we see
the explicit cancellation of the term $\sim  \Dl_T \tl{P}_{qq}(w)$:
\beq \label{coll2}
\Dl_T |M|^2_{\rm coll}
= \fr{4C}{w p_1\C k}  \fr{\Dl_T |M^n_B|^2(t=t')}{C_B C_F}
 \Dl_T P^{<,4}_{qq}(w) + O(\ve^2),
\eeq
for both the anticommuting-$\gm_5$ scheme and for the HVBM scheme.
This demonstrates the required factorization property in the collinear
limit. We also confirm the finding of \cite{V} that,
\beq
\Dl_T P^{<,n}_{qq}(z) =
\Dl_T P^{<,4}_{qq}(z), \SSPP z<1, 
\eeq
in dimensional regularization for the anticommuting-$\gm_5$ 
scheme.  In addition, we have shown that $\Dl_T P^{<,n}_{qq}(z)$ is the 
same in the HVBM scheme. The $\dl$-function part
has an $\ve$-dependence equal to that of the unpolarized
$P^{n}_{qq}(z)$. According to the ``+'' prescription
\cite{AP} one obtains
\beq
\Dl_T P^{n}_{qq}(z) = C_F \lf[ \fr{2}{(1-z)_+} -2 
+\lf( \fr{3}{2} + \fr{\ve}{2} \rt) \dl(1-z) \rt] .
\eeq
As this result agrees with the one obtained as an intermediate step
in the determination of the two-loop $\Dl_T P_{qq}$\cite{V}, in the
anticommuting-$\gm_5$ scheme, which was not
specific to any particular process, this provides a good check
of the process independence of $\Dl_T P^{n}_{qq}(z)$, and hence
of $\Dl_T f_q$.

In order for $\Dl_T f_{q,{\rm DREG}}$ to be meaningful, 
we should be able to relate
it to $\Dl_T f_q$ in some other scheme.
It is easy to see that terms $\sim 1/p_1\C k^{\, 2}$ 
in (\ref{coll}) cannot
arise in four-dimensional schemes with massless quarks -- by four-dimensional,
we mean four-dimensional tensor structure. Terms $\sim 1/p_1\C k^{\, 2}$,
$\sim 1/p_2\C k^{\, 2}$ can only come from $\Dl_T |M_1|^2$, 
$\Dl_T |M_2|^2$ respectively. These squared amplitudes vanish in a 
trivial fashion in four dimensions, so that only $\Dl_T M_1M_2^*$
contributes. In $n$-dimensions this is not true. It is violated by
order $\ve$, due to the relation 
%\beq \label{gammarel}
$\gm_\mu \gm_\rho \gm_\sg \gm^\mu = 4 g_{\rho\sg}I 
- 2\ve \gm_\rho \gm_\sg$.
%\eeq
The most natural four-dimensional alternative to DREG is DRED, where
the tensors and gamma-matrices are kept four-dimensional, but the 
momenta are taken to be $n$-dimensional with, formally, $n<4$. In that
scheme one obtains the collinear behaviour as in (\ref{coll}), but with
$\ve\RA 0$ everywhere, including in the Born term factor. Hence the 
collinear structure is the same as in (\ref{coll2}), 
except with the four-dimensional Born term.
Since DRED and DREG
have the correct factorized form in the collinear limit,
we can use the technique of \cite{BK1} to convert 
subprocess cross sections from
one scheme to another. Differences arising in the subprocess cross
sections are cancelled by  differences in the transversity
distributions, calculable using
the various $\Dl_T P^{n}_{qq}$.
The  conversion
formula is given in \cite{BK1}  and it
has the same form as in the unpolarized and longitudinally polarized
cases (see also \cite{BK2}).

It was checked in \cite{BK2} that the differences in the two-loop
evolutions of the longitudinally polarized parton distributions
could be traced back to the differences in the $n$-dimensional
one-loop longitudinally polarized splitting functions
(as well as the differences in one-loop factorization scheme), 
or equivalently,
the differences in the one-loop parton distributions themselves. 
Since there
are no such differences between the $\Dl_T P^{n}_{qq}$ of the
anticommuting-$\gm_5$ scheme and that of the HVBM scheme
(meaning the transversity distributions are also the same at one-loop), 
we conclude
that the two-loop evolution of $\Dl_T f_q$, and consequently the
corresponding two-loop $\Dl_T P_{qq}$, should be the same in both
schemes as may be checked.

\vsp{-.6cm}

\section{Alternative regularization schemes}

\vsp{-.6cm}
We now consider the details of certain alternatives to DREG.
Although not specifically related to transversity, this is a good 
opportunity to investigate alternatives such as DRED
since DRED is free of the many complications
we had to deal with in DREG in the study of transverse Drell-Yan.
Hence its usefulness may be appreciated here.
As was pointed out in the original one-loop
DRED  calculation of the transverse 
Drell-Yan process \cite{CKM}, a UV counterterm must be added to the 
vertex correction (see also \cite{BK1}).  This is not a problem 
since the counterterm may
be generated by a process independent Feynman rule. Similar terms were 
pointed out in \cite{Kor}. 
Still, it is useful
to consider other alternatives.

For  Drell-Yan (or deep-inelastic scattering)
 the solution is simple.
 One simply uses the
$n$-dimensional metric tensor in the virtual photon propagator. 
Consequently, the $g^{\al\beta}$ in (\ref{trace}) becomes the 
$n$-dimensional one, whereas the $g^{\mu\nu}$ and the gamma
matrices remain four-dimensional.
This projects out only the physical part of the vertex loop. Then, one
also finds the correct behaviour in all the collinear limits, including
for the unpolarized and longitudinally polarized cases and for the
$qg$ subprocess which also arises there -- one finds the
relevant four-dimensional splitting function multiplied by the
$n$-dimensional Born term. 
The $n$-dimensional Born term 
which factors out is the same
as that of the anticommuting-$\gm_5$ scheme of DREG. This is not 
surprising if we note that by taking all metric tensors arising from 
bosons (including $qg$ vertices, etc\ldots)
to be $n(<4)$-dimensional, DRED reduces to DREG (in the absence of 
the $\gm_5$ problem), up to a possible
DRED mathematical inconsistency, which  only rarely arises.  
Hence the Born terms are the same. 
The approach of taking all metric tensors arising from 
bosons to be $n$-dimensional defines a scheme, which  we shall denote
as scheme II (scheme I being DRED with counterterms).
In  scheme II, we observe a  tradeoff between the
ill-definedness of an anticommuting $\gm_5$ (and 
{\em analogous} $\ve^{\mu\nu\rho\sg}$) in $n>4$ dimensions and
that of $g^{\mu\nu}$ in $n<4$ dimensions.
Whatever the approach used, in $n<4$ dimensions
one must first contract all inately
four-dimensional tensors such
as $\ve^{\mu\nu\rho\sg}$ with each other before contracting with 
the $n$-dimensional metric tensor as the contraction of the above
quantities is ill-defined \cite{Sie2}. For situations involving traces
with an odd number of  $\gm_5$'s, or explicit occurrences of
the $\ve$-tensor, scheme II suffers from the same arbitrariness as
the anticommuting-$\gm_5$ scheme of DREG. 
The advantage of anticommuting-$\gm_5$ schemes like scheme II is 
that they generally satisfy Ward identities relevant to electroweak
interactions (and thus require no counterterms) whereas the HVBM
scheme generally requires UV counterterms to restore those identities
as well as various finite renormalizations.

We now return to the first variant on DRED discussed above; the
statements below are not relevant to scheme II.
It is permissible to take the virtual photon as being 
$n$-dimensional, but the radiated gluon must remain four-dimensional
for consistency (i.e.\ same splitting functions and UV sector)
with the usual DRED.
As well, the latter directly leads to the vanishing of $\Dl_T |M_1|^2$, 
$\Dl_T |M_2|^2$ as can be seen from (\ref{trace}). 
The above approach may be stated as a
more general rule for DRED
which {\em so far}
appears to be valid at one-loop\footnote{We have not 
explicitly checked what happens when
3-gluon vertices are present.}: {\em Metric
tensors arising in propagators of bosons not occurring in virtual 
loops and which (unless massive)
never go on shell (where a massless boson might 
develop a mass singularity
or a soft divergence)
are taken as $n$-dimensional}. This saves us from having to add
counterterms in many cases, but not in all cases.
Initial and final state boson lines
are taken as four-dimensional in this scheme. 
Formally, one should add counterterms
in DRED, so in our case
the above procedure  simply amounts to a
trick to avoid that. One {\em can} always add the  counterterms
in this scheme. Then many of them will simply decouple.
If conventional DRED should fail in some 
circumstance, it may prove useful to make the above approach,
which we denote as scheme III, more formal, since it could 
fix  problems which counterterms are not capable of fixing.
One such case which motivates 
further study of scheme III (and also II) is the seeming problem
with conventional DRED pointed out in \cite{Been}. Since that problem
is not in the UV or soft sectors, one does not expect the usual 
counterterm approach to work.

For the longitudinally polarized $qg$ subprocess, in scheme III, all
four-dimensional algebra {\em including} the $\ve$-tensor contraction
was performed first, then the contraction with $g^{\al\beta}$ was
carried out. This simple approach leads to incorrect results in 
scheme II. Additional ad-hoc rules must be introduced. One 
prescription considered
was to first take all the traces in four dimensions
(by using $n$-dimensional metric tensors to project indices
occurring in the traces) and 
contract out all resulting repeated indices, then replace all 
remaining $n$-dimensional metric
tensors with four dimensional ones, contract out the
$\ve$-tensors (so far untouched)
and perform all the remaining contractions in four-dimensions.
Then, the correct form results, with the four-dimensional splitting
function times the $n$-dimensional Born term
arising in the collinear limit. 
The prescription may be stated as
follows: {\em The $\ve$-tensors must be contracted at the end, with only
four-dimensional metric tensors remaining. Those metric tensors are 
obtained by replacing all remaining $n$-dimensional metric
tensors with four dimensional ones (i.e.\ $g_n^{\rho\sg} \RA 
g_4^{\rho\sg}$) after all repeated indices have been contracted out.} 
It should be noted that for
traces involving an even number of $\gm_5$'s, the possible mathematical 
inconsistency is easily removed by performing such traces 
in $n$ dimensions.
Whether $n<4$ or $n>4$ does not matter, as the rules are the same.
Scheme II cannot be considered as DRED or DREG, since the results
obtained correspond to an anticommuting-$\gm_5$ scheme of DREG when
there are an even number of $\gm_5$'s, but DREG gives
no way to define traces with an odd number of $\gm_5$'s and maintain both
cyclicity and anticommutativity of the $\gm_5$.

\section{Summary}

\vsp{-.6cm}

To summarize, we have explicitly demonstrated that DREG leads, in
a non-trivial fashion, to the
correct factorized form for the transverse Drell-Yan squared amplitude
in the limit of collinear gluon radiation. The resulting 
$n$-dimensional one-loop
transversity splitting function is in agreement with
that obtained in \cite{V} as an intermediate step in the determination
of the corresponding two-loop splitting function. The result is the
same in the anticommuting-$\gm_5$ scheme and in the HVBM scheme. 
This implies that the one-loop transversity distributions and their
two-loop evolutions are the same in both schemes.
The transversity distribution which would be obtained  
from the transverse Drell-Yan
process using DREG subprocess cross sections
was shown to be consistent with the corresponding
one of DRED.
In light of the comparative simplicity of the latter scheme, a minor
variant on DRED was considered (scheme III) which can help in
avoiding the addition of the DRED counterterms, at the very
least. The link between DRED and the anticommuting-$\gm_5$ scheme
of DREG was clarified and a specific anticommuting-$\gm_5$
scheme (scheme II) was formulated making use
of it.
 
\vsp{-.6cm}

%\begin{ack}
\section*{Acknowledgements}

\vsp{-.6cm}
The author thanks W.\ Vogelsang for useful correspondence. This work
was supported by U.S.\ Department of Energy contract number
DE-AC02-76CH00016.
%\end{ack}

%\newpage

\twocolumn

%\newpage

\end{document}